\documentclass[twocolumn]{aastex62}
\usepackage{amsmath}
\usepackage{amsfonts}
\usepackage{graphicx}

\allowdisplaybreaks
\graphicspath{{./}{figures/}}

\shorttitle{Binary Magnetosphere of Premerger Neutron Stars}
\shortauthors{J. F. Mahlmann \& A. M. Beloborodov}

\begin{document}

\title{Electrodynamics and Dissipation in the Binary Magnetosphere of Premerger Neutron Stars}

\correspondingauthor{Jens F. Mahlmann}
\email{jens.mahlmann@columbia.edu}

\author[0000-0002-5349-7116]{Jens F. Mahlmann}
\affil{Department of Astronomy \& Astrophysics, Pupin Hall, Columbia University, New York, NY 10027, USA}
\affil{Department of Physics \& Astronomy, Wilder Laboratory, Dartmouth College, Hanover, NH 03755, USA}

\author[0000-0001-7801-0362]{Andrei M. Beloborodov}
\affiliation{Department of Physics, Pupin Hall, Columbia University, New York, NY 10027, USA}
\affiliation{Max Planck Institute for Astrophysics, Garching, 85741, Germany}

\begin{abstract}
We investigate energy release in the interacting magnetospheres of binary neutron stars (BNSs) with global 3D force-free electrodynamics simulations. The system dynamics depend on the inclinations $\chi_1$ and $\chi_2$ of the stars' magnetic dipole moments relative to their orbital angular momentum. The simplest aligned configuration ($\chi_1=\chi_2=0^\circ$) has no magnetic field lines connecting the two stars. Remarkably, it still develops separatrix current sheets warping around each star and a dissipative region at the interface of the two magnetospheres. A Kelvin-Helmholtz (KH)-type instability drives significant dissipation at the magnetospheric interface, generating local Alfv\'enic turbulence and escaping fast magnetosonic waves. Binaries with inclined magnetospheres release energy in two ways: via KH instability at the interface and via magnetic reconnection flares in the twisted flux bundles connecting the companions. Outgoing compressive waves occur in a broad range of BNS parameters, possibly developing shocks and sourcing fast radio bursts. We discuss implications for X-ray and radio precursors of BNS mergers. 
\end{abstract}

\submitjournal{ApJL}


\section{Introduction}

The electromagnetic interaction of two magnetized neutron stars in a tight binary system has been discussed in many previous works \citep[e.g.,][]{Hansen2001,Metzger2016,Lyutikov2018,Wada2020,Most2023}.
In particular, energy dissipation in the binary magnetosphere was considered a source of electromagnetic emission preceding the final merger of the inspiraling companions. Most earlier works discussed how voltage may be induced by the orbital motion of the magnetized stars, leading to ohmic dissipation in the magnetospheric circuit. However, this mechanism is inefficient, because $e^\pm$ pair creation screens the induced voltage and limits ohmic dissipation. Similar screening effects are well known to regulate the low dissipation rates in isolated pulsars \citep{Cerutti2016,Philippov2022REV} and magnetars \citep{Kaspi_2017ARAaA..55..261}. Approximate ``force-free'' models assume completely screened electric fields $E_\parallel$ parallel to the magnetic field $\mathbf{B}$.

A different mechanism can still drive significant dissipation in these systems: the magnetic configuration can become unstable, leading to turbulence excitation and magnetic reconnection. These processes trigger fast energy release, because they develop field variations on very short scales where dissipation becomes inevitable. The magnetospheric dynamics leading to energy release can be simulated using the framework of force-free electrodynamics (FFE) as a first approximation. FFE simulations have previously been performed for orbiting neutron stars \citep{Carrasco2020} and magnetars \citep{Parfrey2013,Carrasco2019,Mahlmann2019,Yuan2022,Mahlmann2023}. Recently, \citet{Most2020,Most2022} used a similar technique to simulate magnetic flares in binary neutron stars (BNSs).

The binary companions are described by their dipole magnetic moments $\boldsymbol{\mu}_1$ and $\boldsymbol{\mu}_2$. Apart from the magnetic field strength, the key parameters are the inclination angles $\chi_1$ and $\chi_2$ of $\boldsymbol{\mu}_1$ and $\boldsymbol{\mu}_2$ relative to the orbital angular momentum of the BNS. \citet{Palenzuela2013} and \citet{Most2020,Most2022} simulated a large set of magnetized BNSs, including anti-aligned binaries with $\boldsymbol{\mu}_1\uparrow\downarrow\boldsymbol{\mu}_2$ ($\chi_1=0^\circ$ and $\chi_2=180^\circ$). In these systems, and more generally in misaligned binaries, a magnetic flux bundle connects the polar regions of the two stars and is continually twisted by their rotation. This leads to quasiperiodic magnetic flares resembling eruptions from the solar corona. \citet{Beloborodov2021} calculated the X-ray emission expected from such flares.

This Letter presents global high-resolution FFE simulations of BNS magnetospheres, designed to analyze the dynamics besides eruptions. We demonstrate the formation of warped current sheets, the Kelvin-Helmholtz (KH) instability at the interface between the companion magnetospheres, and the generation of compressive waves. They are best observed in aligned binaries with $\boldsymbol{\mu}_1=\boldsymbol{\mu}_2$ ($\chi_1=\chi_2=0^\circ$), which have no eruptions but still generate strong dissipation through the KH instability. We also simulate binaries with inclined magnetic dipole moments to study the transition to erupting systems.

\section{Simulation method}
\label{sec:simulations}

We model the two neutron stars as perfect conductors of radius $r_\star$  centered at $\mathbf{x}_{1,2}=(\pm x_0,0,0)$. The simulation is set up in a frame that rotates with the circular orbital motion of the binary with the stars' centers at rest. This frame is noninertial, however, we neglect the Coriolis and centrifugal forces (they would complicate the simulation and only weakly change the binary magnetospheric interaction). Both stars rotate with the same rate $\boldsymbol{\Omega}_{1,2}=\Omega_\star\mathbf{\hat{z}}$ in the simulation frame.\footnote{Our choice of $\Omega_\star$ approximately corresponds to the orbital frequency of the binary $\Omega_0$. When both stars have zero angular momenta in the lab frame (no tidal locking), they are spinning with $-\boldsymbol{\Omega}_0$ in the simulation frame.} We set $x_0/r_\star=3$ (which corresponds to a BNS separation of $6r_\star$) and $\Omega_\star=0.05 c/r_\star$, so the light cylinders of the rotating stars have radius $r_{\rm LC}=20r_\star$. The two stars are endowed with equal dipole moments $\mu_{1,2}$, and their directions are inclined by angles $\chi_{1,2}$ relative to the rotation axis. We have run models with $\chi_1=0^\circ$ and different $\chi_2\in\left\{0^\circ,10^\circ,30^\circ,90^\circ\right\}$. 

Our FFE simulation evolves electric ($\mathbf{E}$) and magnetic ($\mathbf{B}$) fields in time and tracks the magnetospheric charge density $\rho$ and currents $\mathbf{j}$. We integrate Maxwell's equations with corresponding force-free currents while enforcing the constraints
\begin{align}
\rho\mathbf{E}+\mathbf{j}/c\times\mathbf{B}=\mathbf{0}&\qquad\text{(force-free condition)},\label{eq:ffbalance}\\
\mathbf{E}\cdot\mathbf{B}=0&\qquad\text{(ideal fields)}\label{eq:FFI},\\
    E<B&\qquad\text{(magnetic dominance)}\label{eq:FFII}.
\end{align}
We use a high-order finite-volume FFE method and closely follow previous implementations of the rotating pulsar magnetosphere performed with the same code \citep{Mahlmann2019,Mahlmann2020b,Mahlmann2020c,Mahlmann2021} on the infrastructure of the \textsc{Einstein Toolkit}\footnote{\url{http://einsteintoolkit.org/}} \citep{Loeffler2012}. Our numerical framework benefits from the \textsc{Carpet} driver \citep{Goodale2002a,Schnetter2004}. Static mesh refinement of the Cartesian grid with refluxing at internal mesh boundaries\footnote{As provided by Erik Schnetter and David Radice: \url{https://bitbucket.org/dradice/grhydro_refluxing/}} enhances accuracy in the inner magnetosphere. We use a 3D Cartesian grid with cubic cells and simulate a cubic domain of size $L_{0}/r_\star=512$ with a base cell length of $\Delta_{0}/r_\star=4$. Static mesh refinement doubles the resolution consecutively on eight additional levels. The light cylinder scale is captured by a refinement level with $L_6/r_\star=48$ and $\Delta_6/r_\star=1/16$, the stars by a cube of $L_7/r_\star = 10$ and $\Delta_7/r_\star=1/32$, and the interaction region between the stars by a level with $L_8/r_\star = 2$ and $\Delta_8/r_\star=1/64$. We integrate the system using seventh-order spatial reconstruction \citep[MP7,][]{Suresh1997} and a CFL factor of $f_{\rm CFL}=0.2$. Grid cells inside the spherical stars are removed from the active domain. The numerical fluxes on the stars' boundaries match the perfect conductor conditions whenever a directional sweep of the finite-volume integration crosses the stellar boundary \citep{Munz_1999CRASM.328..431}.

\begin{figure}
\centering
\includegraphics[width=0.99\linewidth]{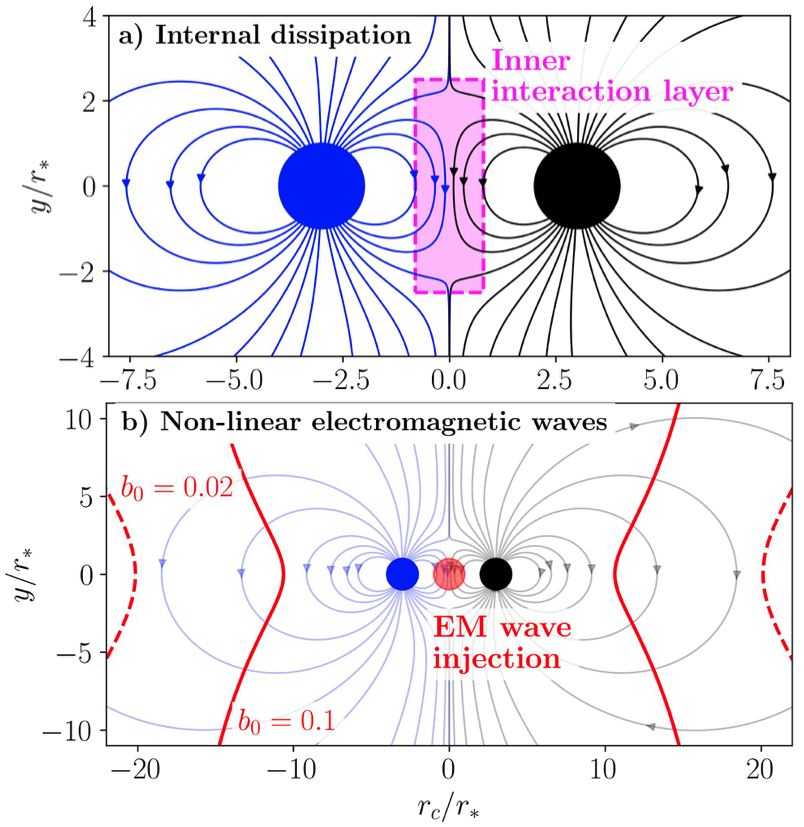}
  \caption{
 Initial (static) state used in the simulation with aligned magnetic moments ($\chi_1=\chi_2=0^\circ$). Top: black and blue curves show the magnetic field lines emerging from the two neutron stars. No
 field lines connect the stars in this configuration. The figure also indicates the energy release region (magenta box) that appears when the stars are spun up and the KH instability develops at the interface of their magnetospheres. Bottom: schematic picture of compressive wave emission from the interface region (red circle). Their relative amplitude $b=\delta B/B$ is $b_0$ in the emission region and grows as the waves propagate to the outer magnetosphere. The red contours show the approximate locations where such waves become nonlinear ($b=\delta B/B_{\rm bg}\sim 1$, leading to shocks), for the initial wave amplitudes $b_0=0.1$ and $b_0=0.02$. The shown magnetosphere is the initial state; the actual configuration changes after spinning up the stars, so the contours of $b=1$ will also change.}
  \label{fig:CFAC}
\end{figure}

\begin{figure*}
\centering
  \includegraphics[width=0.99\linewidth]{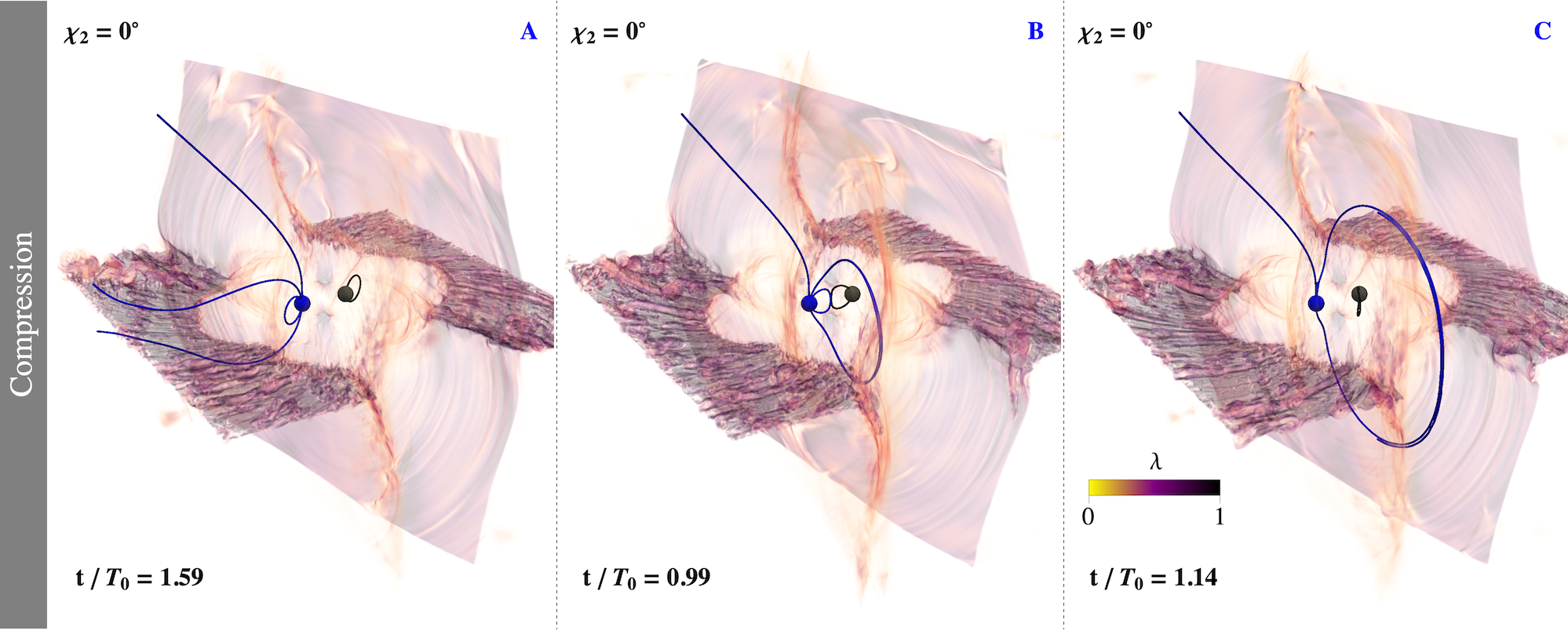}\\
\includegraphics[width=0.99\linewidth]{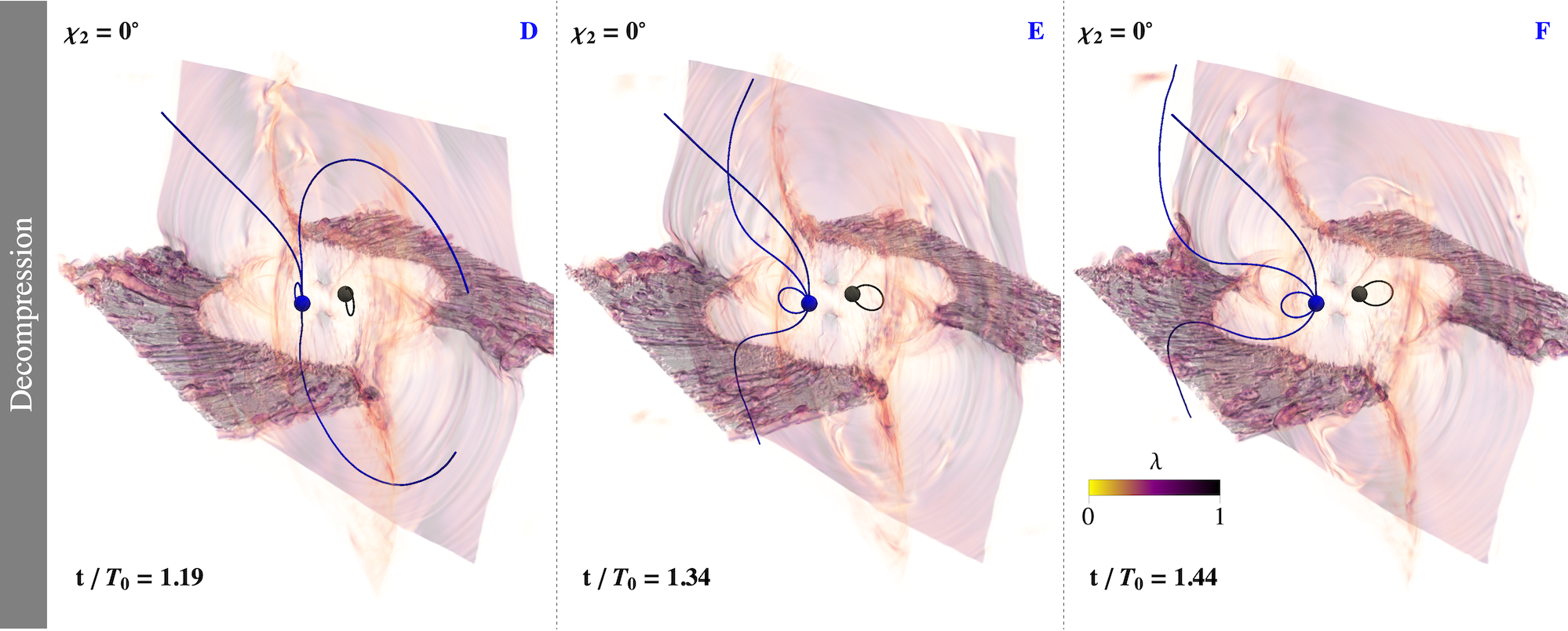}\\
  \caption{Electrodynamics of the aligned binary magnetosphere ($\chi_1=\chi_2=0^\circ$).  The background color shows the field-aligned current density $\lambda$; it sharply peaks at current sheets, which are nearly static, separating domains with different magnetic fluxes. Magnetic field lines are attached to the stars and rotate through the domains. We show three field lines attached to star 1 (blue; footpoints at colatitudes $\theta_{\rm f}=0^\circ$, $15^\circ$, and $30^\circ$) and one field line attached to star 2 (black; footpoint at $\theta_{\rm f}=30^\circ$). The six snapshots are spaced in time and show the evolution during one full rotation cycle. The polar field line ($\theta_{\rm f}=0^\circ$) remains static. The field line with $\theta_{\rm f}=30^\circ$ remains confined in the closed magnetosphere near its star and corotates with moderate deformations. The field line with $\theta_{\rm f}=15^\circ$ experiences strong deformations each rotation cycle. In snapshot \textbf{A}, it is open (its two pieces connecting to infinity are separated by the equatorial current sheet). In snapshot \textbf{B}, it slides off the current sheet and reconnects, forming a closed loop. The rotating star continues to drag the loop in the azimuthal direction. However, it gets temporarily stuck, facing the pressure of the companion magnetosphere that rotates in the opposite direction. Stuck field lines form a domain of compressed magnetic flux accumulated near the vertical midplane of the binary. A warped vertical current sheet separates the compressed domain from the companion magnetosphere, where field lines undergo a decompression phase. As one can see in snapshots \textbf{C} and \textbf{D}, the field line with $\theta_{\rm f}=15^\circ$ inflates in the compressed domain until the loop opens and jumps to the other (far) side of the warped current sheet (snapshot \textbf{E}). There, it catches up with the azimuthal position of its rotating footprint. After decompression, the field line returns toward the equatorial current sheet (\textbf{F}), concluding the cycle. An animated figure version is provided as supplementary material \citep{SupplementaryMediaF}. It shows times $t/T_0=0.80$ to $1.80$. The real-time duration of the animation is 4\,s.}
  \label{fig:figure2a}
\end{figure*}

\section{Aligned magnetic moments}
\label{sec:aligned}

The initial state of the aligned configuration at time $t=0$ in our simulation is shown in Figure~\ref{fig:CFAC}. It has no rotation and is formed by two magnetic dipole moments $\boldsymbol{\mu}_1=\boldsymbol{\mu}_2$ directed along the $z$-axis and separated by distance  $2x_0=6r_\star$. This initial (static) magnetosphere has no electric fields and currents. It is symmetric about the $xy$-, $yz$-, and $xz$-planes. At $t>0$, the two stars are set in rotation with $\boldsymbol{\Omega}_1=\boldsymbol{\Omega}_2$. As discussed below, rotation breaks the symmetry of the magnetic configuration, creates warped current sheets, and triggers energy release in the interface region near the vertical $yz$-plane  of the binary. The energy will be released in two forms, as schematically indicated in Figure~\ref{fig:CFAC}: local turbulent dissipation in the interface region and escaping compressive waves that may form shocks in the outer magnetosphere. In addition, energy will be dissipated in the current sheets. Their shape differs from the current sheets of isolated pulsars, reflecting the different topology of the magnetic flux in a BNS. Below, we first describe the global picture of the rotating BNS magnetosphere and then the energy release due to instabilities at the interface.

\subsection{Global structure of the binary magnetosphere}
\label{sec:topology}

We evolve the BNS magnetospheres for up to two rotation periods $T_0$; the system reaches a quasi-steady state after $t\approx 0.5T_0$. We then examine the topology of the magnetic field lines, identify current sheets, and follow the periodic rotation of individual magnetic field lines attached to the stars to understand the global magnetospheric structure. Figure~\ref{fig:figure2a} presents snapshots of our simulation that visualize the current density\footnote{We plot the quantity $\lambda$ defined by $\boldsymbol{j}_\parallel=\lambda\boldsymbol{B}$, where $\boldsymbol{j}_\parallel$ is the electric current component parallel to $\boldsymbol{B}$. The current in FFE satisfies $\boldsymbol{B}\cdot\nabla\lambda=0$, so $\lambda$ is constant along the magnetic field lines.}
and track the motion of selected magnetic field lines over the full rotation cycle. To clarify the origin of the warped current sheets, it is particularly instructive to examine the evolution of the field line anchored to star 1 at colatitude $\theta_{\rm f}=15^\circ$. The field line forms a large closed loop when approaching the interaction region between the BNS companions. While its footpoints keep rotating, the extended loop becomes stuck when it faces the companion magnetosphere, which occupies the other half-space and rotates with the opposite velocity \citep[see also the animation of Figure~\ref{fig:figure2a};][]{SupplementaryMediaF}. Field lines attached to either companion are squeezed where they approach the interaction region. The accumulating tension eventually inflates and breaks the field lines, allowing them to pass through the middle region and rotate away from the $yz$-plane, with a vigorous decompression.

It is important to note that the reflection symmetries about the $yz$- and $xz$-planes only hold for the static BNS magnetosphere (Figure~\ref{fig:CFAC}). The magnetospheres of rotating BNSs are not symmetric, despite the symmetry in $\boldsymbol{\mu}_1=\boldsymbol{\mu}_2$ and $\boldsymbol{\Omega}_1=\boldsymbol{\Omega}_2$.
The symmetries are broken because the field lines attached to the companions have opposite rotational velocities in the interaction region. Crossing the $xz$-plane then becomes a transition from a compression region (where field lines slow down) to a decompression region, where the broken (open) field lines rapidly leave the interaction zone. The warped separatrix between the magnetic domains of stars 1 and 2 divides the compressing and decompressing regions, which have different magnetic stresses and field directions. This asymmetry implies a jump of the magnetic field across the separatrix, which must be supported by a surface current. This explains why a current sheet separates the BNS magnetospheres.

The warped interface current sheet would not occur in BNS magnetospheres with $\boldsymbol{\mu}_1=\boldsymbol{\mu}_2$ and $\boldsymbol{\Omega}_1=-\boldsymbol{\Omega}_2$, where the rotational velocities would satisfy
reflection symmetries. It would then be sufficient to model a single companion with a mirror boundary condition imposed at the $yz$-plane (Appendix~\ref{app:compressepulsar}). BNSs with mirror symmetry are particularly simple, as they do not develop turbulent dissipation at the interface.\footnote{Mirror symmetry implies vanishing velocity shear at the interface ($\partial_x v_y=0$). 
Then, there is no KH instability and no turbulent dissipation. However, mirror-symmetric BNSs are idealized and not generic. In reality, binary companions likely have magnetic fields of different strengths, $\mu_1\neq\mu_2$, and the magnetospheric interface is shifted toward the star with a weaker field. In this more general case, a strong velocity shear is present at the interface even when $\boldsymbol{\Omega}_1=-\boldsymbol{\Omega}_2$.} 
In this Letter, we focus on configurations with aligned spins $\boldsymbol{\Omega}_1\uparrow\uparrow\boldsymbol{\Omega}_2$ because this case corresponds to a realistic BNS shrinking toward the merger: in
the absence of tidal locking, the stars develop aligned spins when viewed in the frame corotating with orbital motion. In such configurations, a large velocity shear at the interface leads to KH instability and dissipation, as discussed in the next section.

In contrast to single stars, binary magnetospheres have significant open magnetic flux (extending to infinity), even in a static configuration with $\Omega_\star=0$: field lines open up along the interface region because they cannot enter the half-space occupied by the companion. The presence of closed and open magnetic field lines is particularly clear in the $xz$-plane (Figure~\ref{fig:CFAC}). Separatrices between the magnetospheres' open and closed domains intersect the interface between the two magnetospheres. This intersection creates a null point of X-type in the magnetic topology \citep[e.g.,][]{Pontin2022}. Topological singularities 
are a generic feature of binary magnetospheres. In systems with mirror symmetry, the null points are located exactly in the $yz$-plane (Figure~\ref{fig:CFAC}). 

Setting the stars in rotation excites electric currents in the binary magnetosphere, and more magnetic flux opens up. The resulting configuration develops a Y-shaped separatrix (when viewed in the $xz$-plane), a well-known feature of isolated pulsars. The leg of this Y-shape is a current sheet along the equatorial plane, which separates the oppositely directed open fluxes. At the Y-point, the current splits and flows around the closed zone of the magnetosphere. For isolated aligned rotators, the equatorial current sheet beyond the Y-point surrounds the star axisymmetrically. By contrast, in the BNS magnetosphere, the sheet cannot cross the $yz$-plane and instead is bent away from it (Figure~\ref{fig:figure2a}).

We have measured the net Poynting flux $L$ from the binary through spheres of different radii and compared it with the spindown luminosity $L_0$ of a single rotating magnetosphere with $\boldsymbol{\Omega}_1\parallel\boldsymbol{\mu}_1$: $L_0=\mu_1^2\Omega_1^4/c^3$. We have found that $L$ varies between $L\approx 6L_0$ at $r=6r_\star$ and $L\approx 5L_0$ at $r=20r_\star$. The large spindown power reflects the enhanced open magnetic flux that connects the two stars to infinity. A similar effect is also
known (and simplest to calculate) for binaries with mirror symmetry (Appendix~\ref{app:compressepulsar}); in that case, $L\approx 2.5L_0$. 

\begin{figure*}
\centering
 \includegraphics[width=0.99\linewidth]{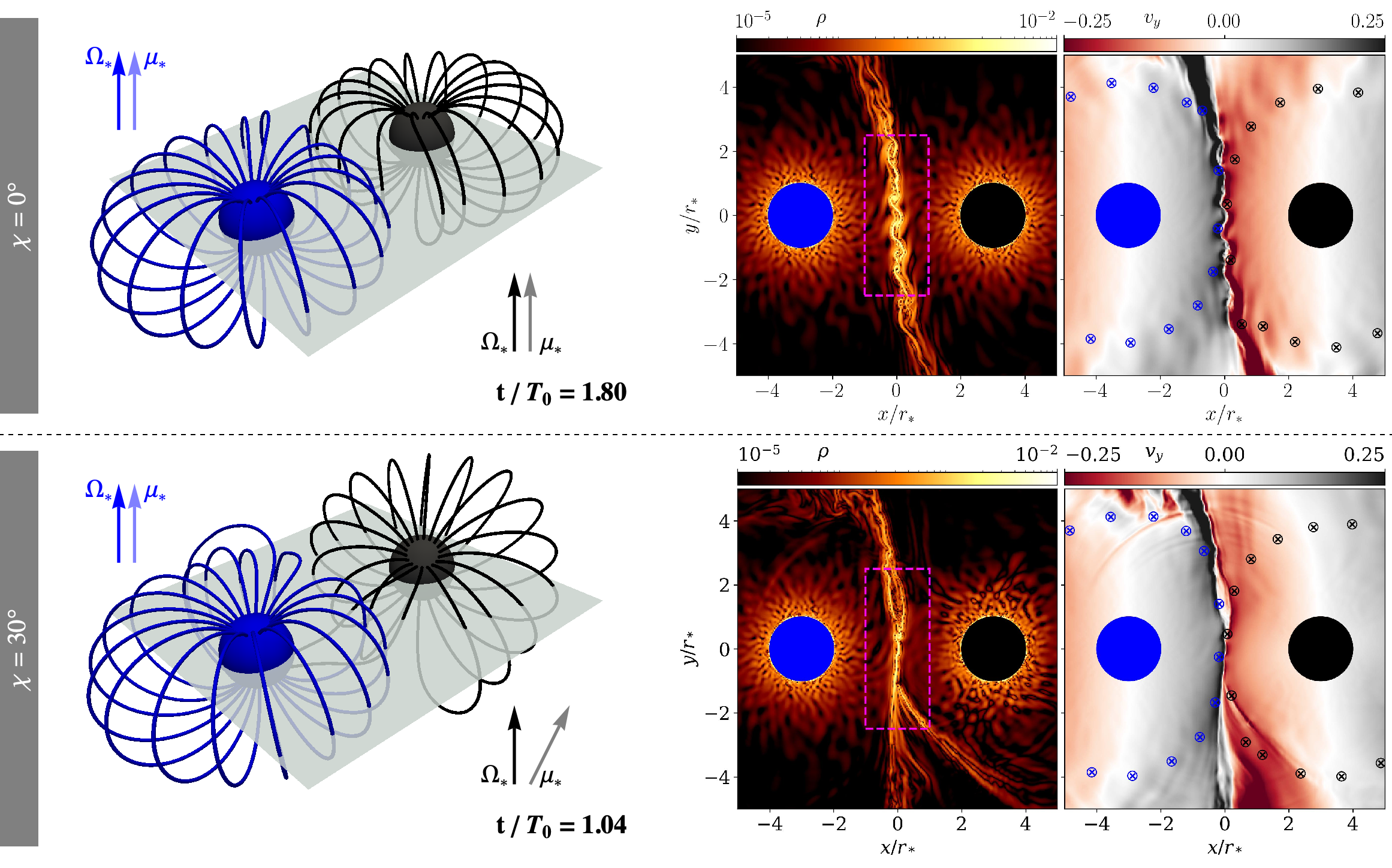}\\
  \vspace{-6pt}
  \caption{
  Snapshots of the simulations with 
  $\chi_1=\chi_2=0^\circ$ (top panels)
and $\chi_1=0^\circ$, $\chi_2=30^\circ$ (bottom panels). Left: selected closed magnetic field lines attached to the rotating stars and interacting in the interface region near the vertical midplane ($yz$-plane). Right: color plots of charge density $\rho=\nabla\cdot\boldsymbol{E}/4\pi$ (related to drift velocity $\boldsymbol{v}/c=\boldsymbol{E}\times\boldsymbol{B}/B^2$) and the velocity component $v_y$, measured in the $xy$-plane. The cross-circles show where the rotating magnetic field lines puncture the plane. The snapshots of $\rho$ and $v_y$ illustrate the motion of the magnetic field lines and show the vortices generated by KH instability in the interface layer (magenta box). The KH vortices create an effective drag between the magnetospheres, intermittently slowing down and releasing the field lines. 
  }
  \label{fig:KH}
\end{figure*}

\subsection{Instabilities, waves, and their dissipation}
\label{sec:instability}

\begin{figure*}
\centering
\includegraphics[width=0.4\linewidth]{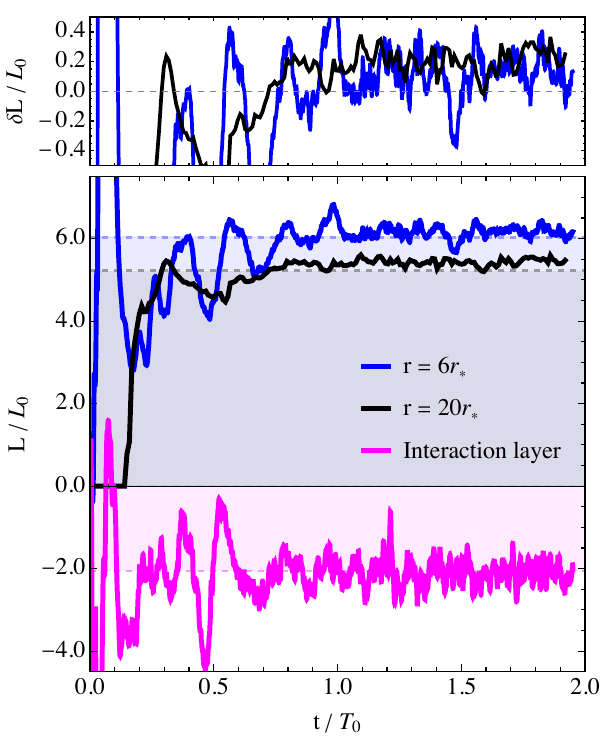}  
    ~
\includegraphics[width=0.5\linewidth]{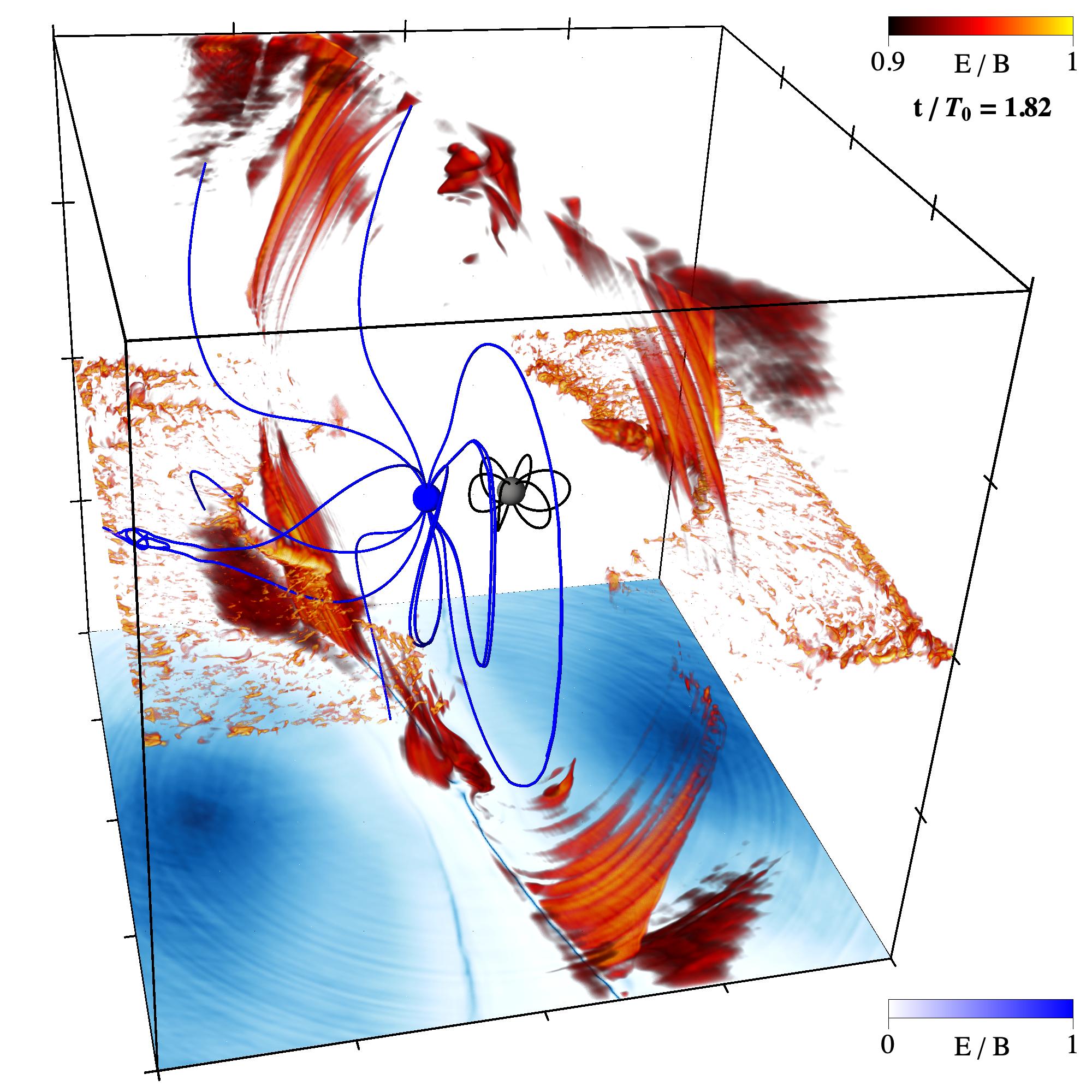}     
  \caption{Energy flows in the aligned binary magnetosphere ($\chi_1=\chi_2=0^\circ$). Left: outgoing Poynting flux through the spheres of $r=6r_\star$ (blue) and $r=20r_\star$ (black) as a function of time. The magenta curve shows the Poynting flux through the boundary of the magenta box indicated in Figure~\ref{fig:KH} ($|x|<0.8 r_\star$, $|y|<2.5r_\star$, $|z|< 2.5r_\star$). The horizontal dashed lines indicate the Poynting fluxes calculated using the time-averaged fields $\mathbf{E}$ and $\mathbf{B}$ (blue/black). We verified that the time-averaged energy flux into the magenta box equals the average dissipation rate in the box (dashed magenta line). We display deviations from the average Poynting fluxes in the top panel. Right: a sample snapshot of the simulation ($t/T_0=1.82$) indicating regions where $E\approx B$. Regions with $0.9<E/B<1$ are shown in volume using red/yellow colors. In addition, $E/B$ is shown on the plane of $z\approx -25$ using blue color. The concentric ripples observed here are a signature of FMS waves, which propagate through the magnetosphere like light in a vacuum. An animated version of the right panel is provided as supplementary material \citep{SupplementaryMediaC} for the time interval $0.93<t/T_0<1.92$; the real-time duration of the animation is 4\,s.}
  \label{fig:PNT_OUTFLOW}
\end{figure*}

We now focus on the energy release at the interface between BNS companions. The rotation of magnetic the field lines attached to stars 1 and 2 implies a large velocity shear ($\partial_x v_y\neq 0$) near the $yz$-plane. In this region, the two rotating magnetospheres experience a KH instability (Appendix~\ref{app:pureKH} discusses its growth rate in FFE). The instability appears as vortices (Figure~\ref{fig:KH}, top right panel), which couple the opposite flows, creating an effective drag between them. The interface instability leads to energy release in the form of turbulence, most of which is dissipated locally, inside the magenta box indicated in Figure~\ref{fig:KH}. The two rotating stars supply power (as Poynting flux) to the interface region, where it is dissipated. Our measurement of the Poynting flux crossing into the magenta box is shown in Figure~\ref{fig:PNT_OUTFLOW}. 

FFE simulations capture dissipation by enforcing the two FFE conditions stated in Equations~(\ref{eq:FFI}) and (\ref{eq:FFII}). We have verified that dissipation due to enforcing $\mathbf{E}\cdot\mathbf{B}=0$ provides the sink of the Poynting flux in the interface region (the magenta curve in Figure~\ref{fig:PNT_OUTFLOW}). In a full plasma simulation, this damping would correspond to the work done by the electric field component $E_\parallel$ parallel to the local magnetic field. Such a damping mechanism is expected in magnetically dominated plasma turbulence. Most of the dissipation executed by the nonideal electric fields is observed in KH vortices. Additional dissipation occurs around the separatrix X-points discussed in section~\ref{sec:topology}.

\citet{Beloborodov2022} proposed KH instabilities in BNS magnetospheres as a source of fast magnetosonic (FMS) waves that may steepen into shocks in the outer magnetosphere. We observe the excitation of FMS waves in the simulation. As expected, they originate in the turbulent interaction region of width $\sim r_\star$ with a modest initial amplitude $b_0=\delta B/B_{\rm bg}$. The relative wave amplitude $b$ grows as they propagate to large radii in the magnetosphere, since $B_{\rm bg}$ decreases faster than $\delta B\propto r^{-1}$. Unlike Alfv\'en waves, FMS waves can propagate across magnetic field lines. They create the impression of concentric waves in $E/B$, resembling those from a stone falling into calm waters (Figure~\ref{fig:PNT_OUTFLOW}, right panel). In most of the domain, the injected FMS waves do not reach nonlinear amplitudes, $b<1$, which implies $b_0<0.1$. However, in some regions, $b$ becomes comparable to unity, indicating $b_0\gtrsim 0.01$.\footnote{For comparison, kink-unstable magnetic flux bundles in a magnetar magnetosphere were found to inject FMS perturbations with $b_0\gtrsim 10^{-3}$ \citep{Mahlmann2023}.} We also observe high-frequency variations in the outgoing Poynting flux from the system, with an amplitude of up to $\delta L/L\approx 5\%$ (Figure~\ref{fig:PNT_OUTFLOW}, top panel). We attribute these variations to outgoing FMS waves generated in the inner interaction region.

Figure~\ref{fig:CFAC} (bottom panel) shows locations in the $xz$-plane of a non-rotating binary where FMS waves with initial $b_0\sim 0.1$ and $b_0\sim 0.02$ would reach $b\sim 1$ and steepen into shocks. Predictions in the figure are only approximate and changed in the full simulation, as rotation changes the magnetosphere at large radii by opening magnetic field lines. In addition, the magnetic configuration looks different near the $yz$-plane, where the two magnetospheres confine and compress each other. Steepening of waves with $b\sim 1$ occurs because $B^2-E^2$ approaches zero \citep{Beloborodov2022}, and this quantity can be directly measured in the simulation. Figure~\ref{fig:PNT_OUTFLOW} shows a 3D simulation snapshot of regions with $E/B>0.9$. FMS waves reach the largest $E/B$ near the $yz$-plane and become capable of launching shocks. 

\section{Misaligned magnetic moments}
\label{sec:resultsinclined}

\begin{figure*}
\centering
\includegraphics[width=0.38\linewidth]{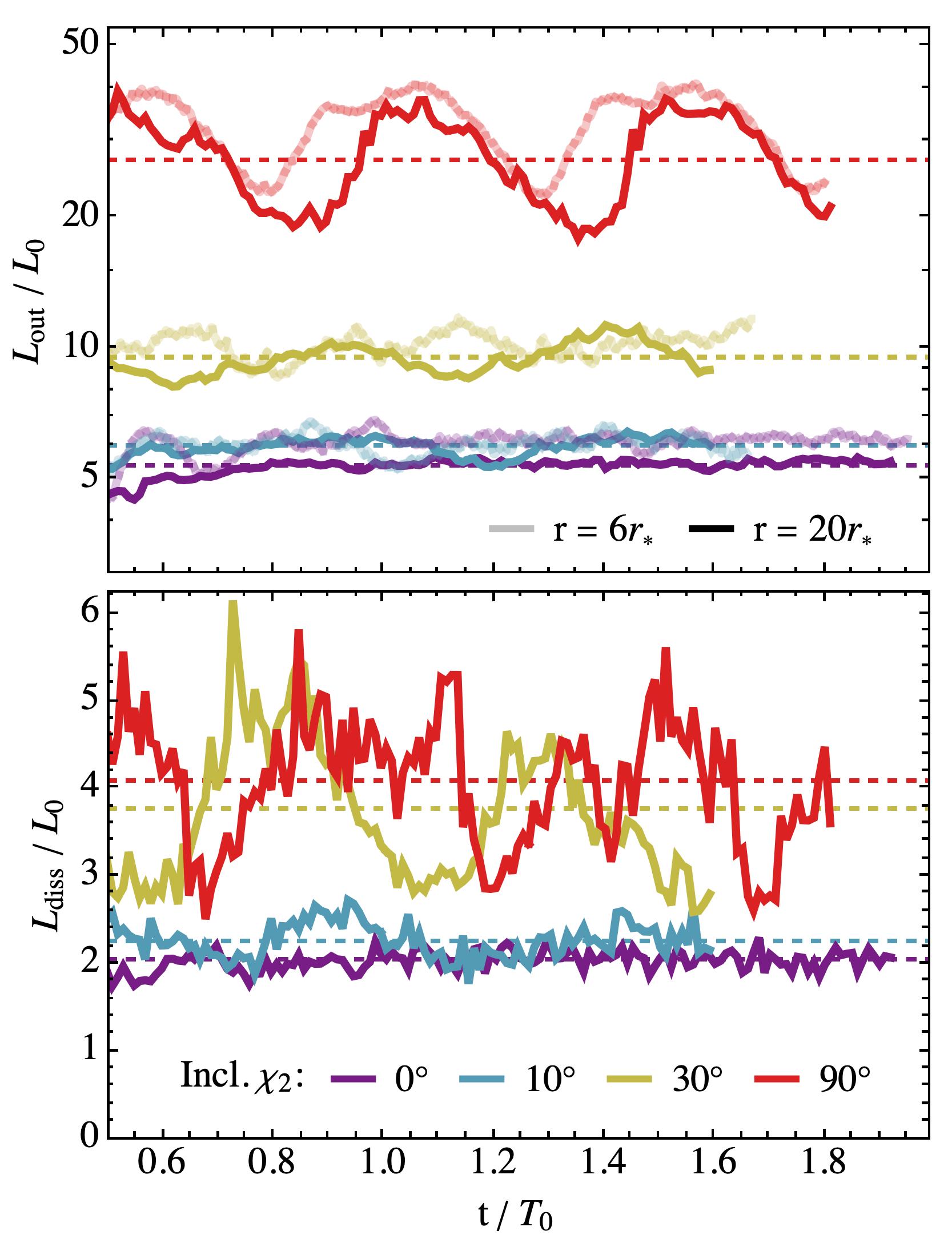}~      
    ~
\includegraphics[width=0.5\linewidth]{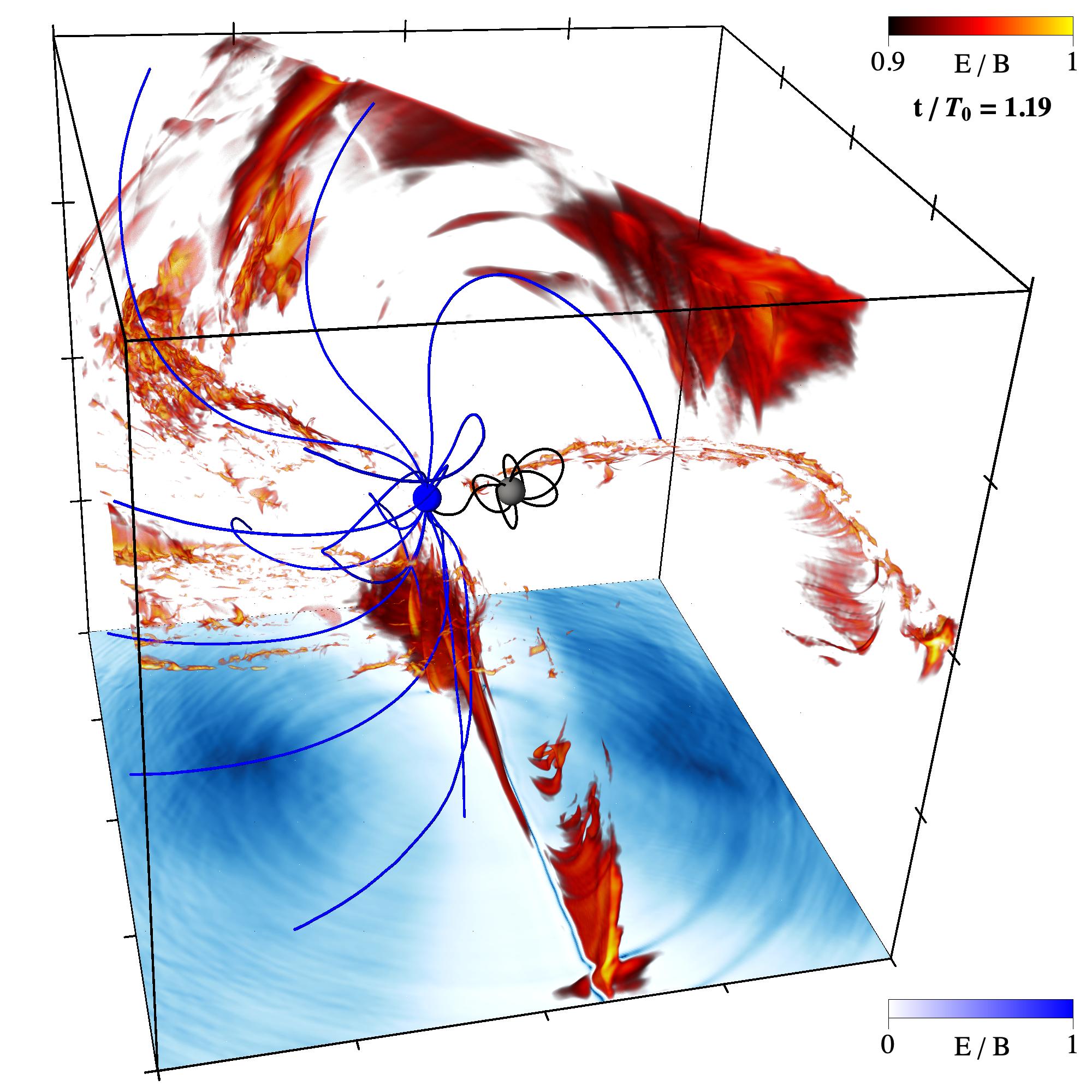}     
  \caption{
  Energy flows of BNS magnetospheres with varying inclinations ($\chi_1=0^\circ$, $\chi_2=0^\circ$, $10^\circ$, $30^\circ$, $90^\circ$). Left: the top panel shows the Poynting flux $L$ through the sphere of
  $r=6r_\star$ (light colors) and $r=20r_\star$ (strong colors). The dashed horizontal lines show the time-averaged luminosities at $r=20r_\star$. The bottom panel displays the dissipation rate driven by enforcing the FFE constraint $\mathbf{E}\cdot\mathbf{B}=0$ measured inside the magenta box indicated in Figure~\ref{fig:KH}. The dashed horizontal lines show the time-averaged dissipation rate. Right: the same as the right panel in Figure~\ref{fig:PNT_OUTFLOW}, but for $\chi_2=30^\circ$. An animated version is provided in the supplementary material \citep{SupplementaryMediaD}. It shows times $t/T_0=0.67$ to $1.66$. The real-time duration of the animation is 4\,s.
  }
  \label{fig:LUMINOSITIES}
\end{figure*}

We have also performed simulations of binaries with $\chi_1=0^\circ$ and $\chi_2=10^\circ,30^\circ,90^\circ$. Their interaction differs from the aligned case in two ways. First, the interface between the misaligned magnetospheres involves a significant magnetic field component parallel to the flow direction (in the $xy$-plane). This field component resists the formation of KH vortices, reducing or (for large inclinations) suppressing the KH instability (Appendix~\ref{app:pureKH}).
This effect is seen in Figure~\ref{fig:KH}, which compares the simulations with $\chi_2=0^\circ$ and $\chi_2=30^\circ$; in the misaligned case, only part of the inner interaction region is filled with KH vortices. Second, in misaligned systems, some magnetic flux connects the companions (Figure~\ref{fig:KH}, bottom panel). A field-line bundle attached to both rotating stars is continually twisted, accumulating energy and intermittently releasing it through instabilities, triggering magnetic flares. Thus, misaligned systems are prolific producers of quasi-periodic eruptions. This process was previously described by \cite{Most2020,Most2022,Most2023}.\footnote{The setup 
of our simulations differs from that in \citet{Most2022}, as we replace the full motion of the binary with two spinning stars in a frame where their centers of mass are at rest (Section~\ref{sec:simulations}) neglecting Coriolis and centrifugal effects. The symmetry of our setup prevents the development of magnetic flares when
$\chi_2=180^\circ$ --- in this special case,
numerical diffusivity helps the system relax to a quasi-equilibrium state. Asymmetric spins could break the symmetry and activate magnetic flares.} The warped current sheets are no longer static 
in misaligned systems; the global magnetic topology shows a complicated quasi-periodic variation.

We analyzed the four simulations with different $\chi_2$ to study how Poynting flux and dissipation rate depend on the misalignment of the two magnetospheres (Figure~\ref{fig:LUMINOSITIES}). We found
that quasiperiodic magnetic eruptions in strongly misaligned binaries dramatically 
enhanced the Poynting flux $L$ (measured at $r=20r_\star$). The time-averaged $L$ is $5.4L_0$ at $\chi_2=0^\circ$, $6.0L_0$ at $\chi_2=10^\circ$, $9.5L_0$ at $\chi_2=30^\circ$, and $27.2L_0$ at $\chi_2=90^\circ$. The time-averaged dissipation rate in the interaction region (the magenta box indicated in Figure~\ref{fig:KH}) is also increased but by a smaller factor: it changes from $2L_0$ to $4L_0$ as $\chi_2$ increases from $0^\circ$ to $90^\circ$. Additional dissipation occurs outside of the interface region enclosed by the magenta box. The contribution of the KH dissipation mode decreases with increasing misalignment; it still accounts for most of the dissipation in the binary with $\chi_2=10^\circ$ and only a fraction at $\chi_2=30^\circ$ and $\chi_2=90^\circ$.

The injection of FMS waves by the interaction of companion magnetospheres remains ubiquitous in misaligned binaries (as evidenced by the concentric features in Figure~\ref{fig:LUMINOSITIES}, right panel). 
Besides the continuous injection of low-amplitude waves and quasiperiodic magnetic eruptions, our simulations show strong wave-like outflows along the extended interaction layer, where regions with $E\approx B$ form intermittently. 

\section{Discussion}
\label{sec:discussion}

Besides eruptions, which are ubiquitous in asymmetric configurations with magnetic field lines connecting the rotating stars \citep{Most2022}, BNSs have another dissipation channel: turbulent heating at the binary interface. For instance, aligned configurations with $\boldsymbol{\mu_1}=\boldsymbol{\mu_2}$ and $\boldsymbol{\Omega_1}=\boldsymbol{\Omega_2}$ maintain a quasi-steady state without eruptions and reveal several remarkable features.

First, the KH instability of the interface generates vortices, leading to turbulent heating. Thus, strong dissipation occurs even without magnetic flux linking the companions. Second, the magnetospheric interaction excites compressive waves, which freely propagate outward and grow in relative amplitude, possibly steepening 
into shocks. Third, the rotation of the magnetic field lines near the interface slows down with strong compression. This nonuniform rotation shapes warped current sheets around the stars. Remarkably, the compression periodically breaks and closes
rotating magnetic field lines, even without magnetic flux connecting the BNS companions.

Magnetic flux linking the companions appears with increasing inclinations of $\boldsymbol{\mu_1}$ or $\boldsymbol{\mu_2}$.
Then magnetic eruptions boost the time-averaged Poynting flux $L_{\rm out}$ emerging from the system (Figure~\ref{fig:LUMINOSITIES}), in agreement with \citet{Most2022}.
At the same time, the dissipation rate in the inner interaction region $L_{\rm diss}$ increases approximately twofold, from $2L_0$ to $4L_0$, where 
\begin{equation}
   L_0\equiv \frac{\mu_\star^2\Omega_\star^4}{c^3}=\frac{B_\star^2 r_\star^6\Omega_\star^4}{c^3},
\end{equation}
with $\mu_\star=|\boldsymbol{\mu}_{1,2}|$, and  $\Omega_\star=|\boldsymbol{\Omega}_{1,2}|$.
For comparison, the power dissipated in an isolated pulsar is $L_{\rm diss}\sim(0.1-0.2)L_0$ \citep{Cerutti2016}. We conclude that dissipation in BNS magnetospheres is an order of magnitude more efficient. 

$L_{\rm diss}$ is the minimum power available for emission from BNSs, and an upper bound is set by the total power extracted from the rotating stars, $L_{\rm tot}=L_{\rm diss}+L_{\rm out}$. The time-averaged Poynting outflow $L_{\rm out}$ reaches $\sim 30 L_0$ in the erupting binaries with large misalignments. Part of $L_{\rm out}$ is carried by an alternating component of $\boldsymbol{B}$ and can be dissipated via magnetic reconnection in the far wind zone \citep{Lyubarsky2001,Drenkhahn2002}, converting to nonthermal emission.

The magnetic fields of old neutron stars are uncertain; strong $B_\star$ are needed for detectable X-ray emission from BNSs, since $L_0\propto B_\star^2$. The power $L_{\rm diss}$ is released in a compact region of size $s$ comparable to a few stellar radii. This corresponds to a large compactness parameter $\ell=\sigma_{\rm T}L_{\rm diss}/s m_ec^3\sim 10^9 L_{\rm diss,44}$, 
where $\sigma_{\rm T}$ is the Thomson cross section. Therefore, $L_{\rm diss}$ immediately converts to X-rays/$\gamma$-rays and $e^\pm$ pairs. \citet{Beloborodov2021} simulated the conversion and the
X-ray spectrum produced by magnetospheric dissipation for $\ell<10^7$. In this regime, radiation leaves the system on a timescale short enough to establish a radiative quasi-steady state. In
more powerful sources, radiation becomes trapped by the opaque $e^\pm$
plasma and is advected out of the system. Full magnetohydrodynamic (MHD) simulations 
will be needed to model the
advection of radiation and its release far from the BNS.

FFE models do not capture the accumulation and advection of dissipated energy. However, they demonstrate the existence of heating mechanisms and their approximate efficiency. The aligned configuration again provides an interesting example. In a quasi-steady state, the heat generated by the KH instability at the interface
must flow out. The periodic opening of the rotating magnetic field lines passing near the interface suggests that the hot MHD outflow will occur mainly in the decompression region, where the open field lines rotate away from the interface. The asymmetric outflow should lead to pulsations of observed emission with a frequency twice the BNS orbital frequency.

The internal shocks suggested by our simulations lead to an interesting possibility of fast radio bursts (FRBs), detectable at relatively low luminosities due to the high sensitivity of radio telescopes. Internal shocks in magnetized $e^\pm$ pair winds were first proposed as an FRB mechanism for magnetars \citep{Beloborodov2017,Beloborodov2020}. The model is based on the synchrotron shock maser, previously investigated for termination shocks of pulsar winds \citep{Hoshino1992,Lyubarsky2014}. Recent kinetic plasma simulations of the shock maser emission are found in \citet{Plotnikov2019} and \citet{Sironi2021}.
\citet{Sridhar2021} argued that the pair-loaded BNS wind may also develop an internal shock (and emit FRBs) due to the steep growth of wind power from the shrinking binary just before merger. 

Our simulations show that the binary magnetosphere is continually filled with kilohertz compressive waves from the unstable interface, and their amplitude can be sufficient to develop shocks. The process of sudden shock formation (by waves reaching amplitudes $\delta B/B_{\rm bg}\sim 1$) has been established by MHD calculations \citep{Beloborodov2022} and plasma kinetic 
simulations \citep{Chen2022,vanthieghem2024}. 
Radiative losses suppress the radio emission from shocks in the magnetosphere. However, when shocks expand to large radii, propagating through the pair-loaded wind, they can emit radio waves with an efficiency up to $10^{-4}$ \citep{Beloborodov2020}. Using $10^{-2}L_0$ as a rough estimate for the shock power in BNS magnetospheres, one could expect a radio luminosity reaching $10^{-6}L_0$. The amplitude and power of compressive waves from the unstable interface grow as the binary shrinks  \citep{Beloborodov2022}. Therefore, a possible radio transient should peak at the final phases of binary evolution. Since its emission is anisotropic, the observed radio signal should be modulated by the BNS orbital motion.

Another mechanism for FRBs from BNSs was proposed by \citet{Most2023}. For some BNS parameters, magnetic eruptions will interact with the outer current sheet around the BNS. This interaction boosts magnetic reconnection, producing small-scale FMS waves, which may later escape as radio waves from the magnetic ejecta
\citep{Lyubarsky2021, Mahlmann2022}.

Features of the FRB mechanisms proposed for isolated magnetars may be shared by BNSs. In particular, the frequency $\nu$ of radio emission from internal shocks scales with the local Larmor frequency in the $e^\pm$ wind and decreases with time proportionally to the local magnetic field in the wind,
$B\propto r^{-1}$ \citep{Beloborodov2017}. In contrast, emission generated by boosted reconnection peaks at a frequency that depends on the magnetic eruption power and implies a correlation between $\nu$ and the luminosity of the generated FRB \citep{Lyubarsky2021}. Both mechanisms produce linearly polarized FRBs.

Future work should overcome the limitations of the presented simulations. Noninertial forces in the simulation frame can be included, similarly to \cite{Most2020}; they become significant with increasing distance from the binary rotation axis and may change the location where FMS waves become nonlinear. More importantly, a full MHD simulation is needed to track the dissipated energy. While FFE demonstrates the development of turbulence and sites of magnetic reconnection, the dissipation processes are not captured in detail. Furthermore, FFE assumes a large magnetization parameter  $\sigma=B^2/4\pi \rho c^2\gg 1$, where $\rho$ is the plasma density. However, dissipation generates thermal energy density $U$, and the effective magnetization is reduced to $\sigma_{\rm hot}\sim B^2/4\pi U\ll \sigma$. It may not far exceed unity, so the electromagnetic field may not strongly dominate the energy of the system. FFE gives a reasonable first approximation when $\sigma_{\rm hot}>1$; however, full MHD simulations should be performed to find $\sigma_{\rm hot}$ and investigate its effects on the magnetospheric dynamics.

\section*{Acknowledgments}

We thank Miguel Á. Aloy, Vittoria Berta, Brian Metzger, Vassilios Mewes, Elias Most, Lorenzo Sironi, Anatoly Spitkovsky, Alexander A. Philippov, Arno Vanthieghem, Yici Zhong, and Muni Zhou for useful discussions. We appreciate the support of DoE Early Career Award DE-SC0023015. This work was supported by a grant from the Simons Foundation (MP-SCMPS-00001470). Our research was facilitated by the Multimessenger Plasma Physics Center (MPPC), NSF grant PHY-2206607. JFM acknowledges support from the National Science Foundation under grant Nos. AST-1909458 and AST-1814708. AMB acknowledges support by NASA grants 80NSSC24K1229, 21-ATP21-0056, 80NSSC24K0282, NSF grant AST- 2408199, and Simons Foundation grant No. 446228. We thank the Max-Planck Institute for Astrophysics (MPA) for hosting a collaborative visit enabling this work. This research was supported by the Munich Institute for Astro-, Particle and BioPhysics (MIAPbP), which is funded by the Deutsche Forschungsgemeinschaft (DFG, German Research Foundation) under Germany's Excellence Strategy: EXC-2094-390783311. Supercomputing resources facilitated the presented simulations: Pleiades (NASA), Ginsburg (Columbia University Information Technology), and Stellar (Princeton Research Computing).

\newpage
\bibliography{literature.bib} 

\appendix

\section{The force-free Kelvin-Helmholtz instability}
\label{app:pureKH}

We follow \citet[][and references therein]{Chow2023,Chow2023b} to derive the dispersion relation for KH modes in relativistic force-free plasmas (Appendix~\ref{app:khlinear}). We then model the instability dynamics in FFE simulations (Appendix~\ref{app:khnumerics}).

\subsection{Linear Analysis}
\label{app:khlinear}

In the plasma rest frame (denoted by a tilde), FFE has two modes with frequency $\tilde{\omega}$ and wavevector $\mathbf{\tilde{k}}$: fast waves ($\tilde{\omega}=\pm c |\mathbf{\tilde{k}}|$) and Alfvén waves  ($\tilde{\omega}=\pm c |\mathbf{\tilde{k}}_\parallel|$). For Alfvén waves, the wavevector is projected along the direction of the background magnetic field (subscript $\parallel$). The dispersion relation for FFE modes is
\begin{align}
    \left(\tilde{\omega}^2-c^2\mathbf{\tilde{k}}^2\right)\left(\tilde{\omega}^2-c^2\mathbf{\tilde{k}}_\parallel^2\right)=0.
\end{align}
We consider a flow with speed $\mathbf{v}_{d\pm}= \pm v_0 \mathbf{\hat{x}}$ that changes direction at a shear layer ($y=0$). The indices plus and minus respectively denote the flow above ($y>0$) and below ($y<0$) the shear layer. For arbitrary wavevectors $\mathbf{k}=(k,l,m)$, we can define the Lorentz invariants 
\begin{align}
    \tilde{\omega}_\pm=\gamma\left(\omega \mp k v_0\right)\qquad\tilde{k}_\pm=\gamma\left(k \mp \omega v_0\right)\label{eq:boostok},
\end{align}
where $\tilde{l}_\pm=l_\pm$ and $\tilde{m}_\pm=m$. For fast waves, we find
\begin{align}
    l_\pm^2=\frac{\gamma^2}{c^2}\left(\omega\mp k v_0\right)^2-\gamma^2\left(k\mp \omega v_0\right)^2-m^2,
\end{align}
from which one can derive the ratio $l_+^2/l_-^2=1$. We analyze perturbations to the electromagnetic energy-momentum flow in the rest frame of the magnetofluid governed by the following equations:
\begin{align}
    \partial_t\mathbf{\tilde{S}}_i+\partial_k \mathbf{\tilde{T}}^k_{~ i}&=0\label{eq:momentum},\\    \partial\mathbf{\tilde{B}}+\nabla\times\mathbf{\tilde{E}} &=0.
    \label{eq:induction}
\end{align}
Here, the assumptions of FFE imply vanishing source terms in Equation~(\ref{eq:momentum}). $\mathbf{\tilde{S}}=\mathbf{\tilde{E}}\times\mathbf{\tilde{B}}$ denotes the Poynting flux, and $\mathbf{\tilde{T}}$ is the electromagnetic stress-energy tensor:
\begin{align}
    \mathbf{\tilde{T}}^k_{~ i}&=-(\tilde{E}^k\tilde{E}_i+\tilde{B}^k\tilde{B}_i)+\frac{1}{2}\left(\mathbf{\tilde{E}}^2+\mathbf{\tilde{B}}^2\right)\delta^k_{~ i}.
\end{align}
The magnetic pressure is defined as $\mathbf{\tilde{p}}_{\rm B}=\mathbf{\tilde{B}}^2/2$. We restrict our analysis to perturbations of the form $\exp{[i(\mathbf{\tilde{k}}\cdot\mathbf{x}-\tilde{\omega}t)]}$ on the background field $\mathbf{B}_0=(\tilde{B}_x,0,\tilde{B}_z)$. In this geometry, the linearized first-order expansion of Equation~(\ref{eq:momentum}) reads
\begin{align}
        -\tilde{B}_x \left(-\tilde{B}_z \delta\tilde{v}_z \tilde{\omega} +\delta\tilde{B}_x \tilde{k}+\delta\tilde{B}_y \tilde{l}+\delta\tilde{B}_z \tilde{m}\right)-\tilde{B}_z \left(\tilde{B}_z \delta\tilde{v}_x \tilde{\omega} +\delta\tilde{B}_x \tilde{m}-\delta\tilde{B}_z \tilde{k}\right)&=0    \\
    \tilde{B}_x \delta\tilde{B}_x \tilde{l}-\tilde{B}_x \delta \tilde{B}_y \tilde{k}-\tilde{B}_z \delta \tilde{B}_y \tilde{m}+\tilde{B}_z \delta\tilde{B}_z \tilde{l}-\delta\tilde{v}_y \tilde{\omega}  \left(\tilde{B}_x^2+\tilde{B}_z^2\right)&=0\\
    -\tilde{B}_z \left(-\tilde{B}_x \delta\tilde{v}_x \tilde{\omega} +\delta\tilde{B}_x \tilde{k}+\delta\tilde{B}_y \tilde{l}+\delta\tilde{B}_z \tilde{m}\right)-\tilde{B}_x \left(\tilde{B}_x \delta\tilde{v}_z \tilde{\omega} -\delta\tilde{B}_x \tilde{m}+\delta\tilde{B}_z \tilde{k}\right)&=0
\end{align}
Similarly, the induction Equation~(\ref{eq:induction}) yields: 
\begin{align}
    \tilde{l}\tilde{B}_x\delta\tilde{v}_y +\tilde{m}\left(\Tilde{B}_x\delta\tilde{v}_z-\tilde{B}_z \delta\tilde{v}_x\right)&=\tilde{\omega} \delta\tilde{B}_x\\
    -\tilde{k}\tilde{B}_x\delta\tilde{v}_y -\tilde{m}\tilde{B}_z \delta\tilde{v}_y &=\tilde{\omega} \delta\tilde{B}_y\\
    \tilde{l} \tilde{B}_z\delta\tilde{v}_y +\tilde{k}\left(\tilde{B}_z\delta\tilde{v}_x-\tilde{B}_x \delta\tilde{v}_z\right)&=\tilde{\omega} \delta\tilde{B}_z
\end{align}
The pressure balance at first order reads $\delta \tilde{p}_B=\tilde{B}_x\delta \tilde{B}_x+\tilde{B}_z\delta \tilde{B}_z$. This combined set of first-order approximations can be cast into an expression for $\tilde{l}$ by eliminating all perturbation amplitudes except $\delta\tilde{v}_y$:
\begin{align}
    \tilde{l}=\frac{2\delta\tilde{v}_y \tilde{p}_B}{\delta \tilde{p}_B\tilde{\omega}}\left[\tilde{\omega}^2 - \left(\tilde{b}_x \tilde{k}+\tilde{b}_z \tilde{m}\right)^2\right]
\end{align}
Here, we use $\mathbf{b}=\mathbf{B}/B_0$. The magnetic pressure is continuous across the shear interface, such that $\tilde{p}_{B+}=\tilde{p}_{B-}$ and $\delta\tilde{p}_{B+}=\delta\tilde{p}_{B-}$. Furthermore, we demand that the displacement $a$ of the shear layer ($y=0$) is continuous along the $y$-direction. From $\gamma \delta v_{y\pm}=\text{d}a/\text{d}t=\partial_t a \pm v \partial_x a=i\left(-\omega\pm v k\right)a$ one finds
\begin{align}
    \frac{\delta v_{y+}}{\gamma\left(v k-\omega\right)}=\frac{\delta v_{y-}}{\gamma\left(v k+\omega\right)}.
\end{align}
Using these substitutions and the Lorentz invariants in Equation~(\ref{eq:boostok}), we can write
\begin{align}
    \frac{l_+^2}{l_-^2}=\frac{(\gamma  (\omega -k v)-\gamma  \tilde{b}_x (k-v \omega )-m \tilde{b}_z)^2 (\gamma  (\omega -k v)+\gamma  \tilde{b}_x (k-v \omega )+m \tilde{b}_z)^2}{(\gamma  (k
   v+\omega )-\gamma  \tilde{b}_x (k+v \omega )-m \tilde{b}_z)^2 (\gamma  (k v+\omega )+\gamma  \tilde{b}_x (k+v \omega )+m \tilde{b}_z)^2}\equiv 1.
\end{align}
We introduce the following substitutions to rewrite the dispersion relation:
\begin{align}
    \phi=\frac{\omega}{\sqrt{k^2+m^2}}\qquad \cos\theta = \frac{k}{\sqrt{k^2+m^2}}
    \label{eq:systemvariables}
\end{align}
Furthermore, $\tilde{b}_x = \cos\Omega$, and $\tilde{b}_z = \sqrt{1-\cos^2\Omega}$. Substituting and solving for the positive imaginary root of $\phi$, we find:
\begin{align}
        \frac{\phi^2}{v^2}&=\frac{\left[\gamma  \cos \theta \left(v-\cos \Omega\right)-\sin \theta \sin \Omega\right]}{\gamma ^2 v^2
   \left(v^2 \cos ^2\Omega-1\right)}\times \left[\sin \theta \sin \Omega+\gamma  \cos \theta \left(v+\cos \Omega\right)\right].
   \label{eq:dispersiongrowth}
\end{align}
For the $\cos\Omega = 0$ limit, where the magnetic background is normal to the flow direction, the instability growth rate is the same as found by \citet[][Equation~38]{Chow2023b},
\begin{align}
    \frac{\phi^2}{v^2}=\frac{\sin^2\theta-v^2}{v^2}.
\end{align}

\subsection{2D FFE Simulations}
\label{app:khnumerics}

\begin{figure*}
\centering
  \includegraphics[width=0.98\linewidth]{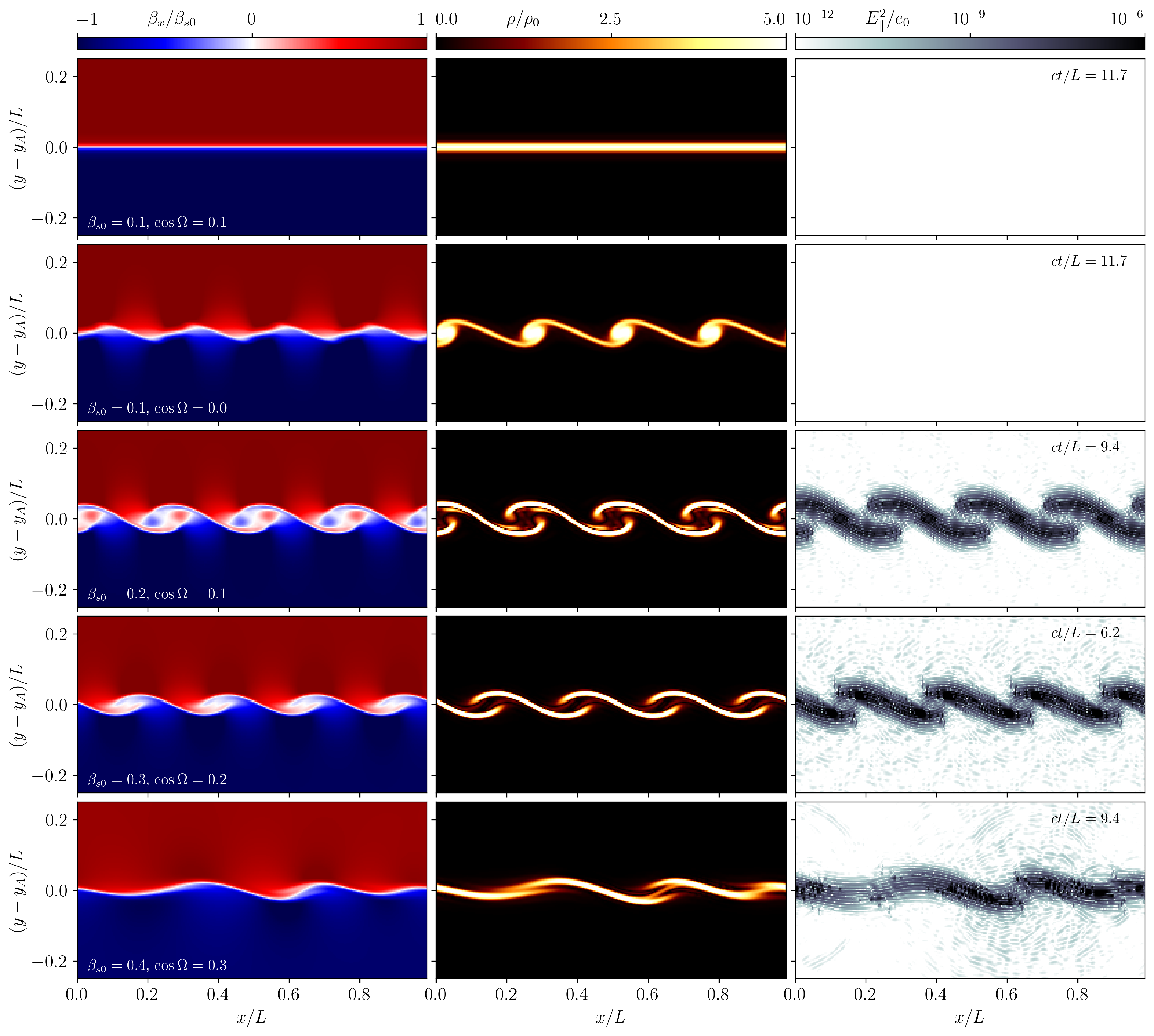}
  \vspace{-11pt}
  \caption{2D shear layers during the linear growth of the force-free KH instability for varying shear velocities $\beta_{s0}$ and magnetic field orientations $\Omega$. We display the normalized drift velocity $\beta_x/\beta_{s0}$ (left), the charge density $\rho/\rho_0$ (middle), and the parallel electric field $E_\parallel^2/e_0$ removed by enforcing the force-free condition $\mathbf{E}\cdot\mathbf{B}=0$ (right). Consistent with Equation~(\ref{eq:dispersion2D}), some configurations will not have any instability (top row). For $\cos\Omega>0$ the instability growth induces parallel electric fields that FFE constraint enforcement removes.}
  \label{fig:FFKH_2D}
\end{figure*}

\begin{figure}
\centering
  \includegraphics[width=0.45\linewidth]{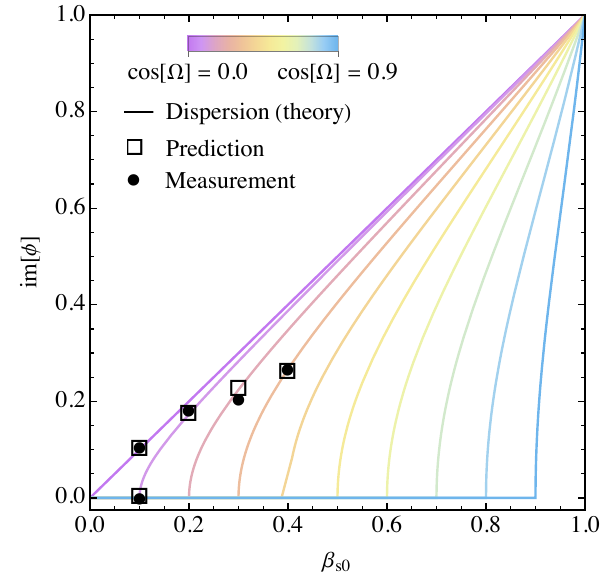}
  \caption{Growth rates of the KH instability as a function of initial shear velocity $\beta_{s0}$: theory (solid lines), predictions for specific simulations (black squares), and simulation measurements (black dots). Solid lines are drawn for different values of $\cos\Omega\in\left\{0,1\right\}$ (see the color-coding). The instability growth is suppressed for $\cos\Omega<\beta_{0}$}
  \label{fig:GROWTH}
\end{figure}

We examine the KH instability in reduced dimensionality (2D), resolving a shear layer and one additional dimension in the $xy$-plane. Then the wavenumber $m$ becomes degenerate; Equation~(\ref{eq:systemvariables}) reduces to $\phi = \omega/|k|$ and $\cos\theta=1$. As a consequence, the growth rate of the KH instability in the force-free  limit (Equation~\ref{eq:dispersiongrowth}) can be written as
\begin{align}
    \frac{\phi^2}{v^2}=\frac{\left(v-\cos\Omega\right)\left(v+\cos\Omega\right)}{v^2\left(v^2\cos^2\Omega-1\right)}.
    \label{eq:dispersion2D}
\end{align}
The instability does not grow if $v\leq\cos\Omega$. In the reduced 2D geometry, the wavenumber $k$ of the surface instability can be inferred by counting the number of KH vortices growing during the linear phase of the instability. We test the performance of the employed FFE method (Section~\ref{sec:simulations}) by comparing the KH growth rates for varying shear velocities $v$ and magnetic background orientations $\Omega$ to the theory outlined in Appendix~\ref{app:khlinear}.

\subsubsection{Setup}

We simulate a periodic double-layered shear flow with asymptotic shear velocity $\beta_{s0}$. The domain is 2D with $x/L\times y/L \in\left[0,1\right]\times\left[0,3\right]$ and resolution $\Delta x/L=\Delta y/L=1/256$. We initialize the magnetic field with $B_{x0} = B_0\cos\Omega$, $B_{y0}=0$, and $B_{z0}=(\gamma_{s0}/\gamma_0)B_0\sin\Omega$. The asymptotic velocity is $\gamma_{0}=(1-\beta_{0}^2)^{-1/2}$, and the local shear velocity is $\beta_{s0}=\beta_{0}\tanh\left[\kappa\left(y-y_A\right)\right]\tanh\left[\kappa\left(y-y_B\right)\right]$, where $y_A$ and $y_B$ define the position of the shear layer and $\kappa=50$. We then prescribe a drift velocity shear $v_{x} = c\beta_{s0}$, and a perturbation $v_{y}=c\beta_{s1}$, with
\begin{align}
    \beta_{s1}&=\chi\beta_{0}\left\{\text{sech}\left[\kappa\left(y-y_A\right)\right]\sin\left[2\pi\kappa x+\phi_A\right]-\text{sech}\left[\kappa\left(y-y_B\right)\right]\sin\left[2\pi\kappa x+\phi_B\right]\right\},
\end{align}
where we use a default choice of $\chi=10^{-4}$ and random phases $\phi_A$ and $\phi_B$. This system is evolved for $ct/L=50$.

\subsubsection{Results}

We display simulation results for different combinations of magnetic field inclination $\Omega$ and shear velocity $\beta_{0}$ in Figure~\ref{fig:FFKH_2D}. As predicted by Equation~(\ref{eq:dispersion2D}), the instability does not grow for $\beta_{0}=\cos\Omega=0.1$ (top panels). Other combinations with $\beta_{0}>\cos\Omega$ show growing instabilities with vortices that resemble characteristic KH eddies, especially in the charge density (middle panels). Then the shear velocity varies around the initial shear layer (shown by different degrees of saturation in the left panels). For disruptions of shear layers with $\cos\Omega>0$, nonideal electric fields parallel to the local magnetic field emerge ($|E_\parallel|>0$). FFE constraint enforcement removes such fields and rapidly dissipates the electromagnetic energy. All the unstable cases in this section rapidly develop KH modes in short times (a few light-crossing times of the shear layer). We quantify the rate of linear instability growth measured in simulations and compare it to the dispersion relation in Figure~\ref{fig:GROWTH}. Our test shows good agreement between simulation and theory, although an accurate determination of the growth rate becomes harder for larger shear velocities $\beta_{s0}$. We conclude that the employed numerical code is capable of resolving the KH instability in FFE with consistent growth rates.

\section{Compressed dipole magnetospheres}
\label{app:compressepulsar}

\begin{figure*}
\centering
  \includegraphics[width=0.98\linewidth]{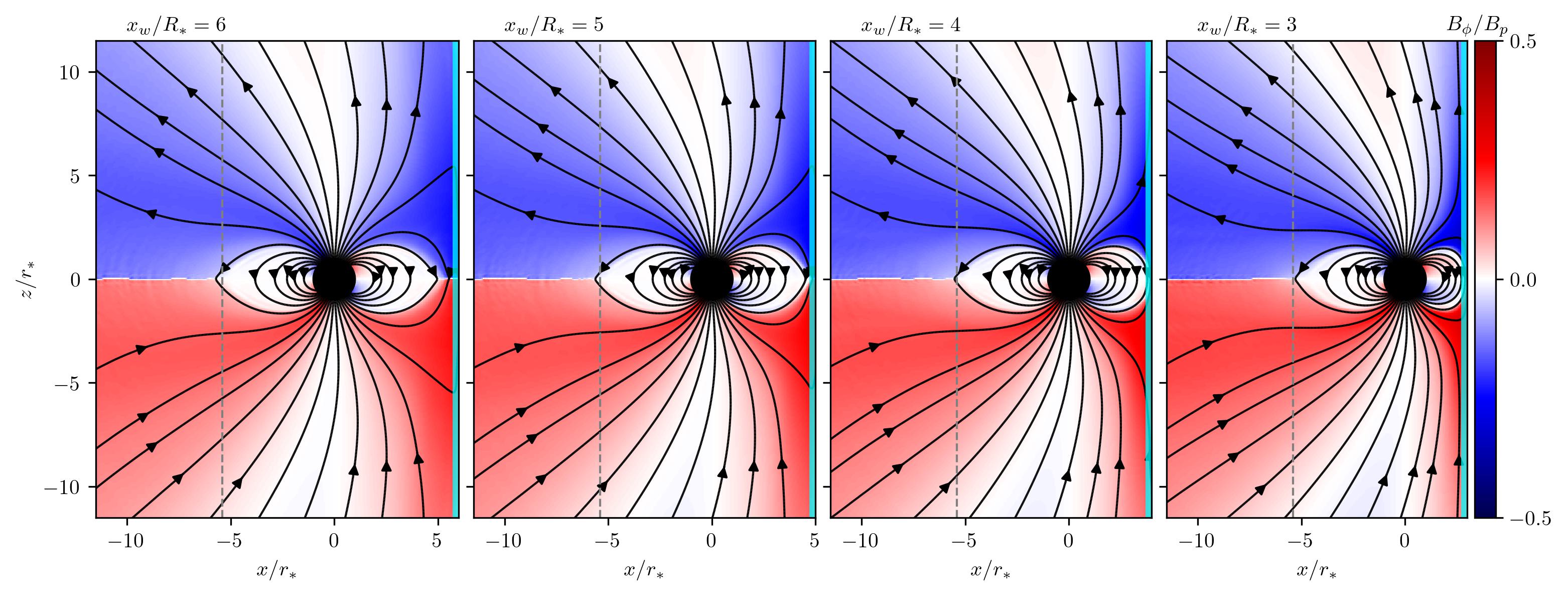}
  \vspace{-11pt}
  \caption{Magnetic fields in 3D simulations of aligned rotator magnetospheres ($\Omega r_*/c\approx 0.2$) next to a perfectly conducting wall at the right end of the domain. The dashed gray lines indicate the nominal light cylinder location. As the conducting wall approaches the star, the magnetosphere is increasingly compressed.}
  \label{fig:DIPWALL_PANELS}
\end{figure*}

\begin{figure}
\centering
  \includegraphics[width=0.45\linewidth]{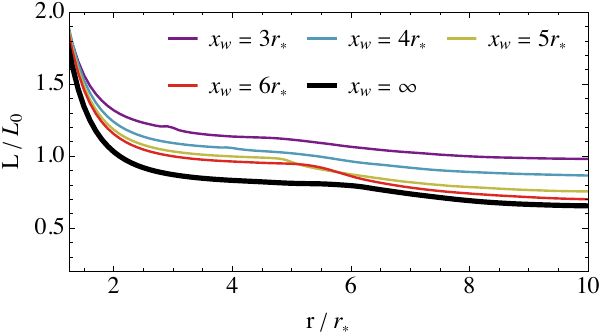}
  \vspace{-11pt}
  \caption{Outgoing luminosity for compressed aligned rotator magnetospheres, normalized to $L_0 = \mu^2\Omega^4/c^3$. Stronger compression drives higher spindown luminosity. The thick black line denotes a reference model without compression.}
  \label{fig:DIPWALL_LUMINOSITY}
\end{figure}

A simplified reference for compressed BNS magnetospheres without interaction can be modeled by placing a perfectly conducting wall next to a neutron star without inclination. We initialize a dipole magnetosphere with rotation, as described in Section~\ref{sec:simulations}. We then add a layer of perfect conductor boundary cells in the $yz$-plane at a distance $x_w$ from the origin. The total domain spans a cube of $x\times y\times z=128r_*\times 128r_*\times 128r_*$ with a base resolution of $\Delta_{xyz}=r_*/2$ ($N_{xyz}=256$ cells). Three additional levels of refinement double the resolution centered around the star, such that $\Delta_{xyz}=r_*/16$ for the innermost cube of $x\times y\times z=32r_*\times 32r_*\times 32r_*$ ($N_{xyz}=512$ cells). We evolve the system for one rotational period of the neutron star. We note that the resolution chosen for the models in this section is low and the duration of the simulations is short, merely serving as a proof of concept.

Figure~\ref{fig:DIPWALL_PANELS} shows the magnetospheric state after one revolution of the star. On the far side (the left part of the panels in Figure~\ref{fig:DIPWALL_PANELS}), we find a typical aligned rotator magnetosphere, with a corotating closed zone extending to the Y-point at the nominal light cylinder radius $r_{\rm LC}=c/\Omega$. The magnetic fields are compressed when reaching the perfectly conducting layer. It compresses the closed zone into the available space between $x\in\left[r_*,x_w\right]$. The closer the wall is to the star, the more field lines are forced to open up when approaching it during their rotation. This is especially noticeable in the right panel of Figure~\ref{fig:DIPWALL_PANELS}, where field lines emerging from the same colatitude on the star extend almost vertically in the right half of the panel while being horizontal on the far side. The increased number of open field lines emerging from the compressed closed zone increases the spindown luminosity. Figure~\ref{fig:DIPWALL_LUMINOSITY} shows the total Poynting flux through spheres at different radii, normalized to $L_0 = \mu^2\Omega^4/c^3$. The Poynting fluxes (measured both in the closed zone, $r<r_{\rm LC}$, and farther away from the star)
increase for smaller separation distances $x_w$. As discussed by \citet{Zhong2024}, the ratio $x_w/r_{\rm LC}$ determines the expected aligned rotator spindown. Here, we probe $x_w/r_\ast\in\left[3,\infty\right]$, corresponding to $x_w/r_{\rm LC}\in\left[0.6,\infty\right]$ and spindowns roughly consistent with \citet[][Figure~3]{Zhong2024}. The setup
of the BNS magnetosphere discussed in the main text uses $x_w/r_{\rm LC}\approx 0.15$, allowing for significantly enhanced spindown luminosities.

For $x_w<r_{\rm LC}$ the Poynting flux decays at distances $r>x_w$, though this decay is less strong than the typical dissipation induced by the magnetospheric current sheet \citep{Spitkovsky2005,Komissarov_2006MNRAS.367...19,Tchekhovskoy2013,Mahlmann2021}. For the shown parameters, the spindown luminosity at large distances differs by a factor of $1.5$ between the simulation with the closest wall ($x_w=3r_*$) and the reference case without any conducting wall in the domain ($x_w=\infty$). Strong Poynting fluxes close to the star also occurred in previous simulations on spherical meshes \citep[see Figure 2 in][]{Mahlmann2021}. The low-order modeling of the spherical stellar surface boundary in a Cartesian mesh likely enhances this effect. We mitigate this numerical noise by higher resolutions around the BNS in the production simulations presented in Section~\ref{sec:simulations}. However, we note that the Poynting fluxes relax to an equilibrium well within the closed zone for distances of $r\gtrsim 2r_*$. Therefore, we conclude that our numerical method resolves compressed BNS magnetospheres with sufficient accuracy. 

\end{document}